\documentclass[epj]{svjour}
\usepackage{graphics}
\usepackage[utf8]{inputenc}
\usepackage{graphicx}
\usepackage{psfrag}
\usepackage{amssymb}
\usepackage{amsmath}
\usepackage{epstopdf}
\usepackage{color}
\usepackage{bm}% bold math
\begin{document}
\title{The Origin of Proton Mass from J/${\rm \Psi}$ Photo-production Data}
\author{
Rong Wang\inst{1,2}\thanks{\emph{Email address:} rwang@impcas.ac.cn}
\and Xurong Chen\inst{1,3}\thanks{\emph{Email address:} xchen@impcas.ac.cn (corresponding author)}
\and Jarah Evslin\inst{1,2}\thanks{\emph{Email address:} jarah@impcas.ac.cn}
}                     % Do not remove
\offprints{}          % Insert a name or remove this line
\institute{Institute of Modern Physics, Chinese Academy of Sciences, Lanzhou 730000, China
\and University of Chinese Academy of Sciences, Beijing 100049, China
\and Guangdong Provincial Key Laboratory of Nuclear Science, Institute of Quantum Matter, South China Normal University, Guangzhou 510006, China
}
\date{Received: date / Revised version: date}
\abstract{
The trace of the stress tensor characterizes the transformation of a theory under rescaling.
In quantum chromodynamics (QCD), this trace contains contributions from the bare masses of the quarks
and also from a purely quantum effect, called the QCD trace anomaly.
It affects all masses in the theory.
We present an estimation of the QCD trace anomaly from the near-threshold J/${\rm \Psi}$ photo-production data
of the GlueX experiment, at JLab.
We apply a vector meson dominance model to describe the photo-production of J/${\rm \Psi}$
and a running strong coupling which includes the nonperturbative effects in the low $\mu^2$ region.
Despite the large uncertainty, we find that the experimental data favors
a small value of the trace anomaly parameter $b=0.07\pm 0.17$.
We report the resulting proton mass decompositions at $\mu^2=0.41$ GeV$^2$ and $\mu^2=4$ GeV$^2$.
 % end of PACS codes
} %end of abstract
\maketitle
%

%%%%%%%%%%% main body %%%%%%%
\section{Introduction}
\label{sec:intro}

Most of the mass of the observable Universe is contained in its protons and neutrons, collectively called nucleons.
However the origin of the nucleon mass is still largely not understood.
The Higgs mechanism provides a relatively modest contribution to the nucleon mass.
The mass of all three valence quarks generated by the Higgs mechanism is much smaller than the mass of the nucleon.
More origins of the nucleon mass is needed.
Nearly all of the rest arises from the various interactions contained in QCD, namely the gluon field.
QCD contains three ingredients.
First, the kinetic energy and self-interactions of gluons are described by Yang-Mills theory \cite{Yang:1954ek}.
Second, the matter is described by quarks \cite{Fritzsch:1973pi}.
Third, with the two ingredients above, the theory leads to divergent results.
To remove these divergences, it must be regularized and renormalized.
In Nobel Prize winning work this was done by 't Hooft and Veltman \cite{tHooft:1972tcz} nearly 50 years ago,
who found that an additional piece, known as the regulator, needs to be added to the theory in the process of removing the divergences.
The final theory is defined by taking a certain limit of the regulator in which it does not vanish,
but rather contributes a finite amount to the various physical quantities.
Its contribution to the trace of the stress tensor is called the trace anomaly.
The stress tensor contains the Hamiltonian, which is the operator which measures the energy of a state
and so determines the rest masses from the QCD dynamics.

According to an analysis of QCD theory,
the mass of a nucleon $M_N$ is decomposed into four terms: the quark energy contribution $M_q$,
the gluon energy contribution $M_g$, the quark mass contribution $M_m$,
and the trace anomaly contribution $M_a$ \cite{Ji:1994av,Ji:1995sv}.
The dependences of these terms on the QCD trace anomaly parameter $b$
and the momentum fraction $a$ carried by all quarks are~\cite{Ji:1995sv},

\begin{equation}
\begin{split}
&M_q=\frac{3}{4}\left(a-\frac{b}{1+\gamma_m}\right)M_N,\\
&M_g=\frac{3}{4}(1-a)M_N,\\
&M_m=\frac{4+\gamma_m}{4(1+\gamma_m)}bM_N,\\
&M_a=\frac{1}{4}(1-b)M_N,
\end{split}
\label{eq:mass-decomposition}
\end{equation}
where $\gamma_m$ is the quark mass anomalous dimension, which can be calculated by perturbative QCD \cite{Baikov:2014qja}.
The magnitude of the anomaly contribution is characterized by the parameter $b$.
Searching a physical interpretation of the mass decomposition, a new type of mass decomposition is suggested \cite{Lorce:2017xzd},
within a semiclassical picture viewing a proton as a hydrodynamical system with pressure.
In the new decomposition, the stability constraint is considered and there are only the internal energy terms of quarks and gluons,
namely $M^{\prime}_q=M_q+M_m$ and $M^{\prime}_g=M_g+M_a$.
The nucleon mass structure is a fundamental question in QCD.

The momentum fraction $a$ carried by the quarks is defined as,
\begin{equation}
a(\mu^2)=\sum_{f} \int_0^1 x[q_f(x,\mu^2) + \bar{q}_f(x,\mu^2)] dx,
\label{eq:mass-fraction-a-def}
\end{equation}
in terms of the proton parton distribution functions $q_f$.
Thanks to decades of the worldwide efforts on both experimental and theoretical sides,
parton distribution functions have been well determined from the global analyses of deep inelastic scattering data.
To complete the proton mass decomposition,
only the trace anomaly parameter $b$ remains poorly constrained by the experimental data.

Although the proton mass is mainly generated from the dynamical chiral symmetry breaking,
a small contribution does arise from the masses of the quarks.
The quark mass contribution is determined by a sum of the scalar charges of the nucleon,
which is related to the QCD trace anomaly paramter $b$ via
\begin{equation}
\begin{split}
bM_N&=\left<N|m_u \bar{u}u+m_d\bar{d}d|N\right> + \left<N|m_s \bar{s}s |N\right> \\
    &=m_l\left<N|\bar{u}u+\bar{d}d|N\right> + m_s\left<N|\bar{s}s |N\right> \\
    &=\Sigma_{\pi N} + \Sigma_{sN}.
\end{split}
\label{eq:quark-mass-term-decomposition}
\end{equation}
The scalar charges of the nucleon are vital in searching for a weakly-interacting massive particle,
one type of the dark matter candidates, since they are the important parameters in the calculation
of the scattering between the nucleon and the dark matter particle \cite{Goodman:1984dc,Akerib:2016vxi,Aprile:2017iyp}.
The scalar nucleon matrix element of up and down quarks, $\Sigma_{\pi N}$, is about $45$ MeV
as has been determined from the low energy $\pi-N$ scattering amplitude \cite{Gasser:1990ce,RuizdeElvira:2017stg,Ling:2017jyz}
and with more data from the pionic-atom spectroscopy \cite{Alarcon:2011zs,Alarcon:2012nr}.
Less is known about the strange term $\Sigma_{sN}$.
It is believed to be large as the mass of the strange quark is large.
However, recent Lattice QCD calculations find that $\Sigma_{sN}$ is comparable to $\Sigma_{\pi N}$.
In particular, the QCDSF Collaboration finds $\Sigma_{sN} = 11\pm 13$ MeV \cite{Bali:2011ks}
and the $\chi$QCD Collaboration finds $\Sigma_{sN} = 40\pm 12$ MeV \cite{Yang:2015uis}.
The strange $\Sigma$ term is also suggested to be small around 16 MeV from an effective field theory \cite{Alarcon:2012nr}.

Low energy scattering between heavy quarkonium and nucleon can be used to probe
the properties of the nucleon. It is not difficult to compute the scattering amplitude using
the operator production expansion. In the quarkonium rest frame, if the incident nucleon energy
is much smaller than the binding energy of the quark-antiquark pair of the quarkonium,
then the leading twist gluon field operator alone already provides a good approximation \cite{Kharzeev:1995ij}.
In the non-relativistic domain, the heavy quarkonium is mainly sensitive to the chromo-electric part
of the gluon field of the nucleon. This is because the velocity of the heavy quark inside of the heavy quarkonium is small
and the chromo-magnetic part is suppressed by powers of velocity.
Hence the heavy quarkonium-nucleon scattering amplitude is determined by the strength of the nucleon color gauge field \cite{Kharzeev:1995ij},
which is the main contribution to the nucleon mass.

Y. Hatta, A. Rajan, and D.-L. Yang had tried to extract the trace anomaly based on a holographic QCD
framework, from the J/${\rm \Psi}$ photo-production data near threshold \cite{Hatta:2018ina,Hatta:2019lxo}.
From the analysis, they pointed out that the $t$-dependence is important for the extraction of the parameter $b$,
and the $t$-dependence is argued to be complicated. Within the holographic model, they find a hint that
the parameter b is small and the gluons play an more important role as the origin of the nucleon mass.
However their result shows that the trace anomaly is loosely constrained by the current data.
It is worthwhile and necessary to look for other reliable and well-acknowledged theories
to extract the QCD trace anomaly.
In this work we will extract the QCD trace anomaly parameter $b$ from the scattering amplitude between charmonium
and nucleon, using the theoretical framework \cite{Kharzeev:1995ij,Kharzeev:1998bz} based on a vector meson dominance (VMD) model.

The QCD trace anomaly parameter $b$ is an essential ingredient for the nucleon mass decomposition,
and for the strange $\Sigma$ term.
The aim of this work is to report a new and relatively powerful experimental constraint on $b$
and to describe how it affects our picture of the origin of the proton's mass.

\section{Forward cross-section of J/${\rm \Psi}$ photo-production off the proton}
\label{sec:data-analysis}

\begin{figure}[htp]
\centering
\includegraphics[width=0.48\textwidth]{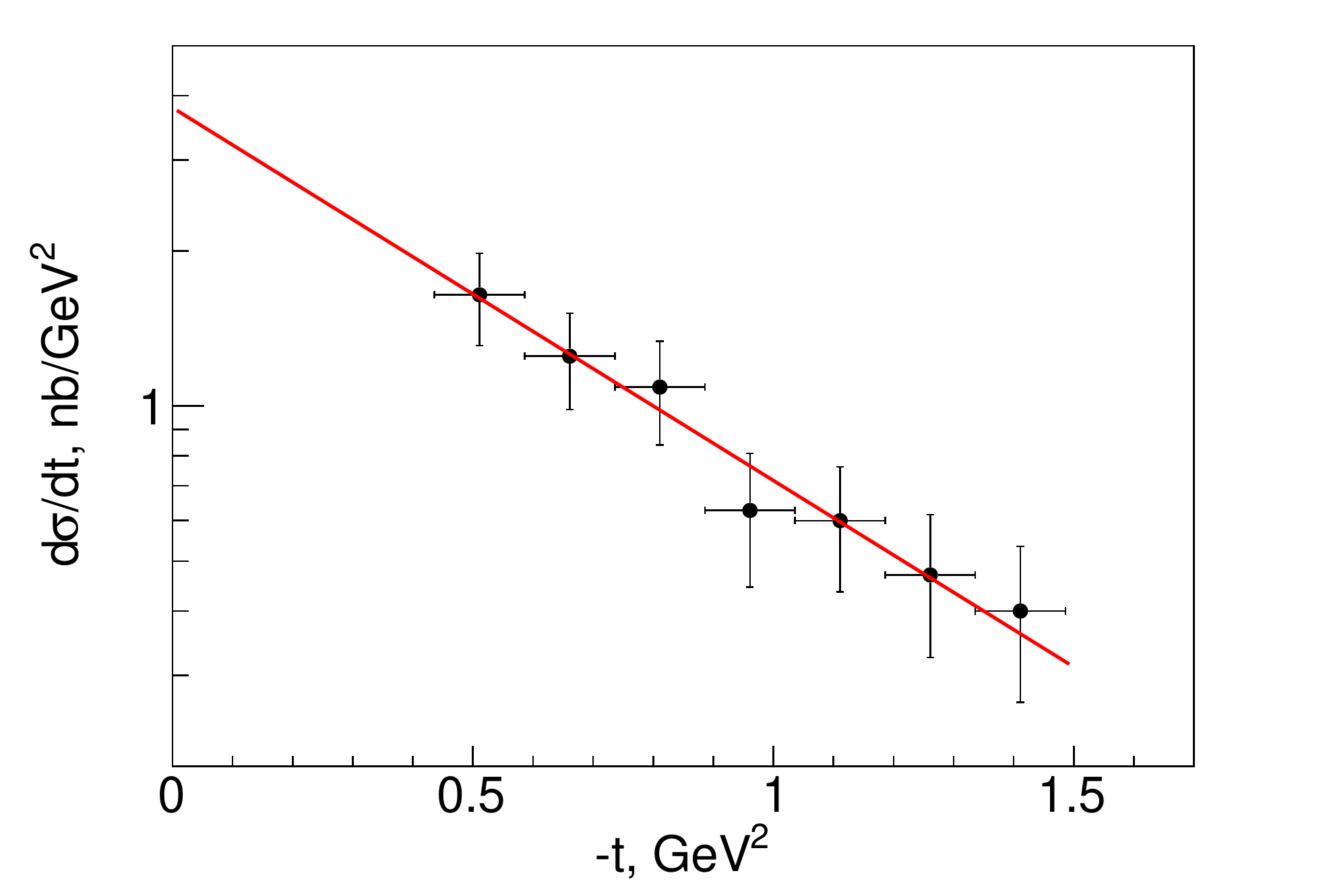}
\caption{
The differential cross section of J/${\rm \Psi}$ photo-production
near threshold as measured by the GlueX Collaboration \cite{Ali:2019lzf}.
Only statistical uncertainties are shown.
}
\label{fig:JPsi-cross-section-GlueX}
\end{figure}

We will perform our analysis using the differential cross-section data recently published
by the GlueX Collaboration at Jefferson Laboratory (JLab) \cite{Ali:2019lzf}.
The cross-section data is on the exclusive J/${\rm \Psi}$ photo-production
from a bremsstrahlung photon beam of 10.72 GeV energy on average.
At this energy, $t_{min}$ of the exclusive reaction is -0.4361 GeV$^2$.
The differential cross-section as a function of $-t$ is shown in Fig. \ref{fig:JPsi-cross-section-GlueX}.
The data is fit with an exponential function ($d\sigma/dt=d\sigma/dt|_{t=0}\times e^{-kt}$)
which describes the $t$-dependence of the cross-section.
$d\sigma/dt|_{t=0}$ is found to be $3.8\pm 1.4$ nb/GeV$^2$,
and the exponential slope $k$ is found to be $-1.67\pm 0.38$ GeV$^{-2}$.
Errors are determined by fixing $\chi^2 - \chi^2|_{\text{best fit}}=1$.

The dependence of the forward cross-section $d\sigma/dt|_{t=0}$ on the QCD trace anomaly parameter $b$
is discussed in Sec. \ref{sec:VMD-model}.
Comparing the measured forward cross-section to the $b$-dependent prediction yields a fit for $b$.

\section{VMD model and J/${\rm \Psi}$ photo-production}
\label{sec:VMD-model}

In the VMD model, the forward cross-section of J/${\rm \Psi}$ photo-production on the nucleon is formulated as \cite{Kharzeev:1998bz},
\begin{equation}
\begin{split}
&\frac{d\sigma_{\gamma N\to J/\Psi N}}{dt}\bigg|_{t=0} =\\
&\frac{3\Gamma(J/\Psi\to e^+e^-)}{\alpha m_{J/\psi}}\left(\frac{k_{J/\Psi N}}{k_{\gamma N}}\right)^2
 \frac{d\sigma_{J/\Psi N\to J/\Psi N}}{dt}\bigg|_{t=0},
\end{split}
\label{eq:VMD-model}
\end{equation}
where $k_{ab}^2=[s-(m_a+m_b)^2][s-(m_a-m_b)^2]/4s$ denotes the squared center-of-mass momentum
of the corresponding two-body reaction, and $\Gamma$ stands for the partial decay width of J/${\rm \Psi}$.
The center-of-mass energy $\sqrt{s}$ is 4.58 GeV with the incident photon of 10.72 GeV.
The decay width of J/${\rm \Psi}$ to electron-positron pair is 5.547 keV \cite{Tanabashi:2018oca}.
$\alpha=1/137$ is the fine structure constant.

The differential cross-section of J/${\rm \Psi}$-N interaction is given by,
\begin{equation}
\frac{d\sigma_{J/\Psi N\to J/\Psi N}}{dt}\bigg|_{t=0} = \frac{1}{64\pi}\frac{1}{m_{J/\Psi}^2(\lambda^2-m_N^2)}\left|F_{J/\Psi N}\right|^2,
\end{equation}
where $F_{J/\Psi N}$ denotes the invariant J/${\rm \Psi}$-N scattering amplitude,
and $\lambda=(p_{N}p_{J/\Psi}/m_{J/\Psi})$ is the nucleon energy
in the charmonium rest frame \cite{Kharzeev:1998bz}.
At low energies, the amplitude takes the form \cite{Kharzeev:1995ij},
\begin{equation}
\begin{split}
F_{J/\Psi N} &\simeq r_0^3 d_2 \frac{2\pi^2}{27}\left(2M_N^2 -\left<N\bigg|\sum_{i=u,d,s}m_i\bar{q}_i q_i\bigg|N\right>\right)\\
&\simeq r_0^3 d_2 \frac{2\pi^2}{27}\left(2M_N^2 -2bM_N^2\right)\\
&\simeq r_0^3 d_2 \frac{2\pi^2}{27}2M_N^2 (1 - b).
\end{split}
\label{eq:MPsiN-modeling}
\end{equation}
The sum of the $\Sigma$ terms of the quarks is directly connected to the trace anomaly parameter $b$.
Note that Kharzeev uses a relativistic normalization of the hadron state $\left<N|N\right>=2M_NV$,
where $V$ is a normalization volume \cite{Kharzeev:1995ij}.
In the chiral limit and low energy scattering, the mass of a hadron state comes purely from the quantum fluctuations of gluons.
Since we do not know how much the quark mass contribute to the trace of the energy-momentum tensor,
the factor $(1-b)$ is introduced here to represent the unknown size of the trace anomaly contribution.
The ``Bohr" radius $r_0$ of the charmonium in Eq. (\ref{eq:MPsiN-modeling}) is given by,
\begin{equation}
r_0=\left(\frac{4}{3\alpha_s}\right)\frac{1}{m_c}.
\end{equation}
The Wilson coefficient $d_2$ in Eq. (\ref{eq:MPsiN-modeling}) is calculated as \cite{Peskin:1979va,Kharzeev:1995ij,Kharzeev:1996tw},
\begin{equation}
d_n^{(1S)}=\left(\frac{32}{N_c}\right)^2 \sqrt{\pi} \frac{\Gamma(n+\frac{5}{2})}{\Gamma(n+5)},
\end{equation}
where $N_c$ is the number of colors.
The renormalization scale $\mu^2$ is taken to be the ``Rydberg" energy squared $\epsilon_0^2$
of the bound state of the heavy quark-antiquark pair \cite{Kharzeev:1995ij,Kharzeev:1996tw}.

\section{The running strong coupling}
\label{sec:QCD}

\begin{figure}[htp]
\centering
\includegraphics[width=0.48\textwidth]{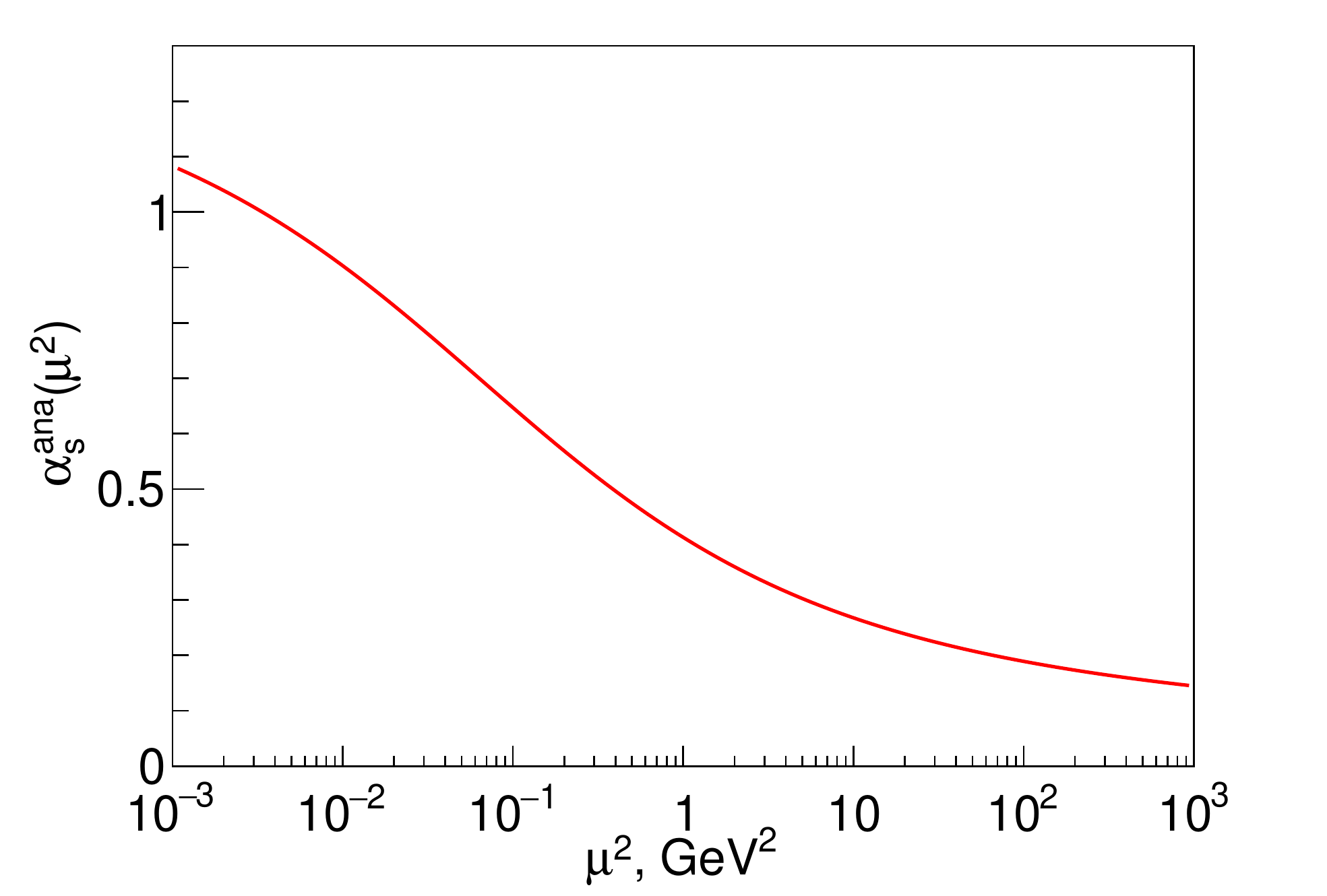}
\caption{
The strong coupling constant $\alpha_s$ in the analytic approach with $\Lambda^2=0.0643$ GeV$^2$.
}
\label{fig:alphas-analytic}
\end{figure}

To avoid the Landau pole of the running strong coupling constant at low $\mu^2$,
we fix $\alpha_s$ using the analytic approach of Ref.~\cite{Deur:2016tte},
in which QCD nonperturbative effects are folded into the coupling.
The resulting analytic expression for $\alpha_s$ is
\begin{equation}
\alpha_s^{\text{ana}}=\frac{4\pi}{\beta_0}\left(\frac{1}{ln(Q^2/\Lambda^2)} + \frac{\Lambda^2}{\Lambda^2-Q^2} \right).
\label{eq:alphas-ana}
\end{equation}
The term $\Lambda^2/(\Lambda^2-Q^2)$ is a nonperturbative power law contribution,
which cancels the Landau pole at $\Lambda^2$.

The QCD running coupling can be determined from a global analysis of the deep inelastic scattering data
in a wide $Q^2$ range, such as GRV98 \cite{Gluck:1998xa} and CT14 \cite{Dulat:2015mca}.
One finds $\Lambda^2=0.0643$ GeV$^2$ in Eq. (\ref{eq:alphas-ana}) by matching the CT14(LO)'s $\alpha_s$ at
the charm quark threshold $m_c^2=1.69$ GeV$^2$. Using the CT14(LO) analysis, $\alpha_s$ is 0.3719 at 1.69 GeV$^2$.
The analytic coupling $\alpha_s^{\text{ana}}=0.3612$ also matches the GRV98(LO)'s $\alpha_s=0.3626$,
at the chosen charm quark threshold of 1.96 GeV$^2$ according to GRV.
Fig. \ref{fig:alphas-analytic} shows the running strong coupling $\alpha_s$ as a function of the energy scale $\mu^2$.

\section{Trace anomaly and proton mass decomposition}
\label{sec:trace-anomaly}

The strong coupling in the calculation depends on the energy scale of ``Rydberg" energy $\epsilon_0$.
Thinking about pulling apart a $c\bar{c}$ pair to generate a $D\bar{D}$ pair,
a naive estimate of the ``Rydberg" energy $\epsilon_0$ is $m_{D}+m_{\bar{D}}-m_{J/\Psi}$ \cite{Kharzeev:1995ij}.
Using the masses of the D meson and the J/${\rm \Psi}$ meson, $\epsilon_{\text{0}}^{2}$ is calculated to be 0.41 GeV$^2$.
Around a low energy scale of 0.41 GeV$^2$, the charm quark mass is near the pole mass of 1.67 GeV \cite{Tanabashi:2018oca}.
At the ``Rydberg" energy, the parameters for the VMD model are determined as following:
$\alpha_s=0.494$ and $r_0=0.318$ fm.

\begin{figure}[htp]
\centering
\includegraphics[width=0.45\textwidth]{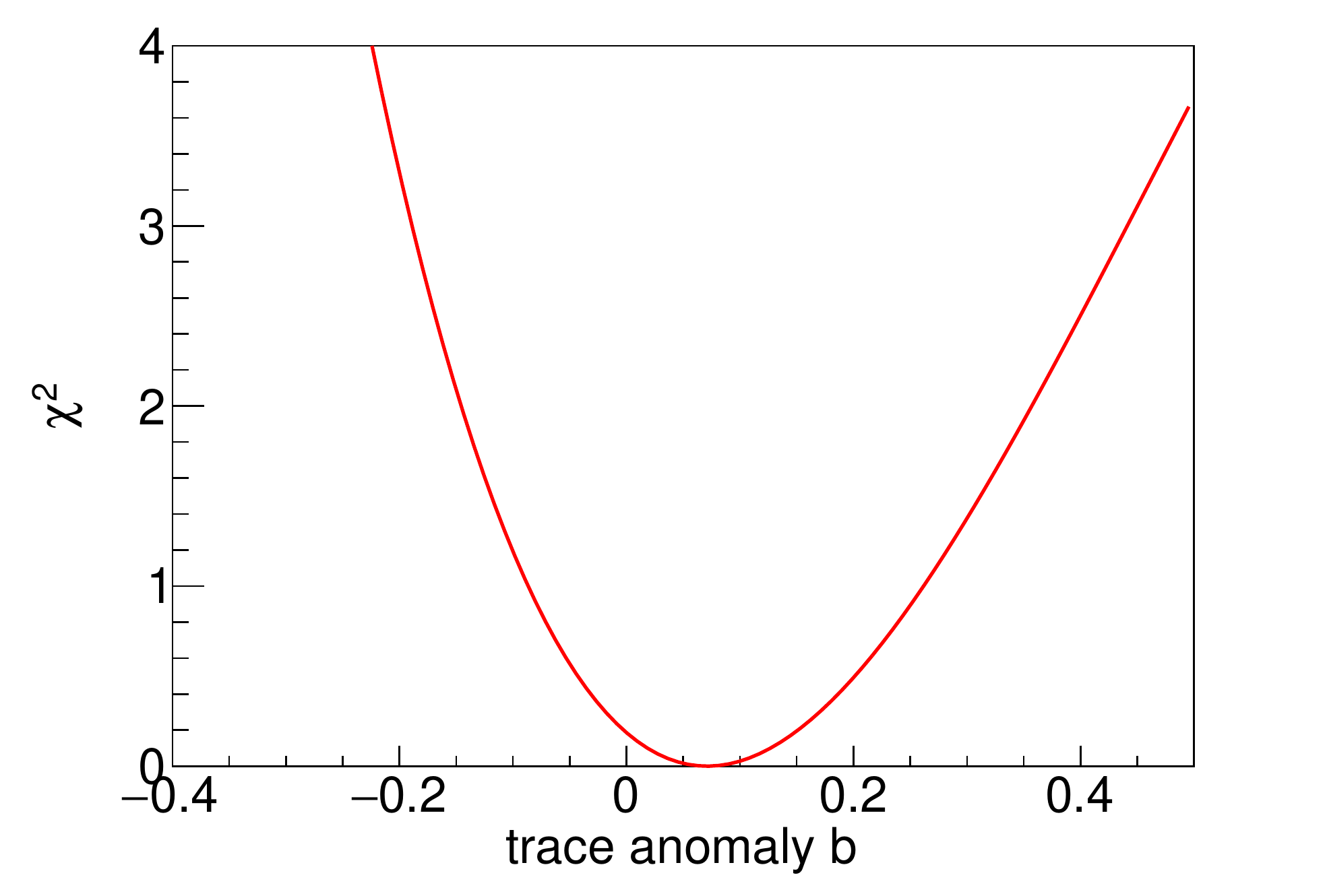}
\caption{
$\chi^2$ as a function of the QCD trace anomaly parameter $b$ under a VMD model description.
In the VMD model, $\epsilon_0^2=0.41$ GeV$^2$, $\alpha_s(\epsilon_0^2)=0.494$, and
$m_c(\epsilon_0^2)=1.67$ GeV are applied as the inputs.}
\label{fig:chi2-vs-b}
\end{figure}

Applying Kharzeev's formulism, the QCD trace anomaly parameter is obtained to be $b=0.07\pm 0.17$.
The $\chi^2=\left(d\sigma/dt|_{t=0}^{\text{exp}}-d\sigma/dt|_{t=0}^{\text{VMD model}}\right)^2/\left(\delta_{d\sigma/dt|_{t=0}}^{\text{exp}}\right)^2$
as a function of $b$ is shown in Fig. \ref{fig:chi2-vs-b}.
If we have more forward cross-section data at several photon energies near the threshold,
the combined uncertainty of the extracted trace anomaly can be reduced.

\begin{figure}[htp]
\centering
\includegraphics[width=0.32\textwidth]{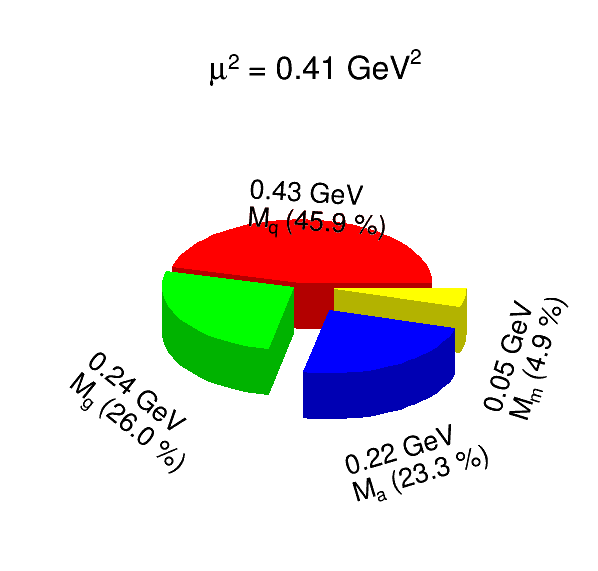}
\includegraphics[width=0.32\textwidth]{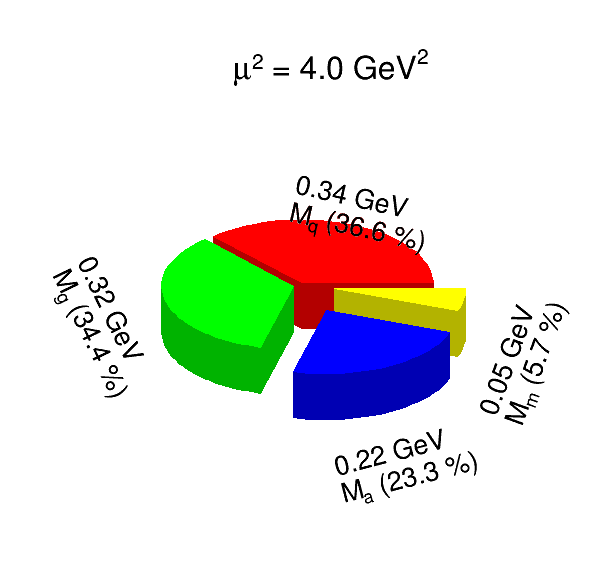}
\caption{
Decompositions of the proton mass at $\mu^2=0.41$ GeV$^2$ and $\mu^2=4$ GeV$^2$, with $b=0.07$.}
\label{fig:proton-mass-pie-charts}
\end{figure}

With the dynamical parton distribution functions generated from the DGLAP equation with parton-parton
recombination corrections \cite{Wang:2016sfq}, we calculate the momentum fraction $a$ carried by all quarks
at the ``Rydberg" energy $\epsilon^2_0$ of J/${\rm \Psi}$ and at $\mu^2=4$ GeV$^2$,
which are 0.654 and 0.541 respectively.
In principle if the trace anomaly parameter $b$ is scale-invariant,
we can obtain the proton mass decomposition at any scale $\mu^2$.
The proton mass decompositions at $\mu^2=0.41$ GeV$^2$ and $\mu^2=4$ GeV$^2$
are shown in Fig. \ref{fig:proton-mass-pie-charts} for illustrations.
In our calculations, the quark mass anomalous dimensions at 0.41 GeV$^2$ and 4 GeV$^2$
were evaluated to be 0.691 and 0.315, respectively \cite{Baikov:2014qja}.
Note that the anomalous dimension $\gamma_{\text{m}}$ defined by Ji \cite{Ji:1994av,Ji:1995sv,Hatta:2019lxo,Hatta:2018sqd} is
$\gamma_{\text{m}}=\mu \frac{d\text{ln}(Z_{\text{m}})}{d\mu}=2\mu^2 \frac{d\text{ln}(Z_{\text{m}})}{d\mu^2} = -2\mu^2 \frac{d\text{ln}(m_{\text{R}})}{d\mu^2}$,
which is twice of the minus of the definition in Ref. \cite{Baikov:2014qja}.
A Lattice QCD calculation gives $M_q=(0.33\pm 0.04)M_N$, $M_g=(0.37\pm 0.05)M_N$,
$M_a=(0.23\pm 0.01)M_N$, and $M_m=(0.09\pm 0.02)M_N$ at $\mu^2=4$ GeV$^2$ \cite{Yang:2018nqn}.
Our result on the proton mass decomposition is close to the result from the Lattice QCD simulation.

\section{Discussions and summary}
\label{sec:discussions}

With the QCD trace anomaly parameter $b=0.07$ extracted from the current data
on the near-threshold J/${\rm \Psi}$ photo-production and the $\Sigma_{\pi N}$ term
determined from the experimental measurements of $\pi$-N scattering \cite{Gasser:1990ce},
the strange $\Sigma$ term is estimated to be around 21 MeV, which is quite small.
Nonetheless, the obtained $\Sigma_{sN}$ term is more or less consistent with the results
of Lattice QCD ($11\pm 13$ MeV and $40\pm 12$ MeV) \cite{Bali:2011ks,Yang:2015uis} and the phenomenological result \cite{Alarcon:2012nr}.
From our analysis, the sum of the $\Sigma$ terms is $bM_N=66$ MeV.
This lies between the Lattice QCD predictions $49\pm 25$ MeV \cite{Bali:2011ks} and $86\pm 19$ MeV \cite{Yang:2015uis}
from two independent collaborations.
More precise data on the near-threshold heavy quarkonium photo-production are needed.

The trace anomaly parameter $b$ is very sensitive to the parameter $r_0$,
based on the VMD model adopted in this analysis.
In theory, the ``Bohr" radius $r_0$ depends on the two key QCD inputs
-- the strong coupling constant and the charm quark mass.
Therefore we should look for more experimental constraints
on the ``Bohr" radius $r_0$ of the heavy quark-antiquark pair, the strong coupling $\alpha_s$,
and the charm quark mass $m_c$. Higher-twist calculations and the systematic uncertainty of the model
should be investigated as well. So far, the uncertainty of the trace anomaly extracted in this work
is quite large. More statistics are needed on the experimental side for the photo-production
of J/${\rm \Psi}$ near threshold. Moreover, our analysis also could be tested with the near-threshold $\Upsilon$(1S)
production using a quasi-real photon at a low energy electron-ion collider \cite{Chen:2018wyz}.
Since the ``Rydberg" energy $\epsilon_0$ of $\Upsilon$(1S) is higher than that of J/${\rm \Psi}$,
Eq. (\ref{eq:MPsiN-modeling}) is valid for a wider kinematical range near the threshold of $\Upsilon$(1S) production.
The theoretical uncertainty of the running strong coupling is much smaller at a higher energy scale $\epsilon_0$.
The chromo-magnetic contribution to the $\Upsilon$(1S)-N interaction is even smaller
as the velocity of the bottom quark inside $\Upsilon$(1S) is even smaller,
compared with the J/${\rm \Psi}$-N interaction. Hence the theoretical framework in this work is more
suitable for the near-threshold $\Upsilon$(1S) photo-production.

In summary, we have extracted the QCD trace anomaly parameter $b$ from recently published JLab data.
This small value of $b$ implies that the QCD trace anomaly is definitely big for the proton state.
We have provided a new proton mass decomposition and suggested a small strange $\Sigma_{sN}$ term.

\section*{Acknowledgments}
This work is supported by the Strategic Priority Research Program of Chinese Academy of Sciences under the Grant NO. XDB34030301.
JE is supported by the CAS Key Research Program of Frontier Sciences grant QYZDY-SSW-SLH006 and the NSFC MianShang grants 11875296 and 11675223.
He also thanks the Recruitment Program of High-end Foreign Experts for support.

\bibliographystyle{unsrt}
\bibliography{refs}{}

\end{document}